\documentclass[iop]{emulateapj}
\usepackage{amsmath}
\usepackage{epstopdf}
\usepackage{multirow}
\usepackage{subfigure}
\usepackage{longtable}

\newcommand{\rearth}{R$_{\oplus}$}
\newcommand{\msun}{M$_{\odot}$}

\newcommand{\mps}{m~s$^{-1}$}
\newcommand{\teff}{$T_{\rm eff}$}
\newcommand{\kep}{{\it Kepler}}

\slugcomment{Submitted to {\it The Astrophysical Journal}}

\shorttitle{M Dwarf Metallicities and Giant Planet Occurrence}
\shortauthors{Gaidos \& Mann}

\begin{document}


\title{M Dwarf Metallicities and Giant Planet Occurrence:\\ Ironing
  Out Uncertainties and Systematics}

\author{Eric Gaidos \altaffilmark{1}} 
\affil{Department of Geology \& Geophysics, University of Hawai`i at M\={a}noa, Honolulu, HI 96822}
\email{gaidos@hawaii.edu}
\and
\author{Andrew W. Mann \altaffilmark{1,2}}
\affil{Department of Astronomy, University of Texas at Austin, Austin, TX 78712}

\altaffiltext{1}{Visiting Astronomer at the Infrared Telescope Facility, which is operated by the University of Hawaii under Cooperative Agreement no. NNX-08AE38A with the National Aeronautics and Space Administration, Science Mission Directorate, Planetary Astronomy Program.}
\altaffiltext{2}{Harlan J. Smith Postdoctoral Fellow}

\begin{abstract}
Comparisons between the planet populations around solar-type stars and
those orbiting M dwarfs shed light on the possible dependence of
planet formation and evolution on stellar mass.  However, such
analyses must control for other factors, i.e. metallicity, a stellar
parameter which strongly influences the occurrence of gas giant
planets.  We obtained infrared spectra of 121 M dwarfs stars monitored
by the California Planet Search (CPS) and determined metallicities
with an accuracy of 0.08 dex.  The mean and standard deviation of the
sample is -0.05 and 0.20 dex, respectively.  We parameterized the
metallicity dependence of the occurrence of giant planets on orbits
with period less than 2~yr around solar-type stars and applied this to
our M dwarf sample to estimate the expected number of giant planets.
The number of detected planets (3) is lower than the predicted number
(6.4) but the difference is not very significant (12\% probability of
finding as many or fewer planets).  The three M dwarf planet hosts are
not especially metal rich and the most likely value of the power-law
index relating planet occurrence to metallicity is 1.06 dex per dex
for M dwarfs compared to 1.80 for solar-type stars; this difference,
however, is comparable to uncertainties.  Giant planet occurrence
around both types of stars allows, but does not necessarily require,
mass dependence of $\sim 1$ dex per dex.  The actual
planet-mass-metallicity relation may be complex and elucidating it
will require larger surveys like those to be conducted by ground-based
infrared spectrographs and the {\it Gaia} space astrometry mission.
\end{abstract}
\keywords{stars:abundances -- stars:low-mass -- stars:fundamental parameters -- planets and satellites:gaseous planets -- planets and satellites:formation -- methods:spectroscopic}

\section{Introduction}

The formation of gas giant planets like Jupiter is a central problem
in planetary science.  In the prevailing scenario, a solid core of
rock and ice is the nucleation site for the runaway accretion of a
gaseous envelope \citep{Helled2014}.  This model predicts that
planet-forming disks with more condensible elements form cores more
readily and that metal-rich stars are more likely to host giant
planets.  This correlation has been confirmed by Doppler radial
velocity (RV) surveys of solar-type stars
\citep{Gonzalez1998,Fischer2005} and all evidence points to a similar
correlation for M dwarfs \citep{Johnson2009,Neves2013,Mann2013b}.

Stellar mass may also be an important determinant of giant planet
formation.  If more massive stars are born with more massive disks,
something suggested but not unambiguously supported by observations
\citep{Andrews2013}, giant planet occurrence should also increase with
stellar mass.  \citet{Cumming2008} analyzed Doppler radial velocity
detections in the California Planet Search (CPS) and estimated that M
dwarfs have 3-10 times fewer giant planets within 2.5~AU than
solar-type stars.  \citet{Johnson2010} re-visited the stellar mass and
metallicity distributions of CPS-detected giant planets around both
solar-mass and M dwarf stars and found that: (i) all M dwarf giant
planet hosts are metal-rich, and (ii) accounting for metallicity, M
dwarfs host about half as many giant planets as do solar-mass stars.
Conversely they found that giant planets are found even more
frequently around evolved stars that derive from more massive
progenitors.  

More recently, \citet{Montet2014} used adaptive optics imaging to rule
out stellar companions to M dwarfs in cases where a long-term drift in
radial velocity was found by Doppler observations.  This allowed them
to better constrain the occurrence of giant planets at separations to
$\sim$20~AU from 111 M dwarfs monitored by the CPS.  Combining this
with survey data for higher mass stars, they found a strong
metallicity dependence ($3.8 \pm 1.2$ dex per dex) but a much weaker
and less significant dependence on mass ($0.8\pm1$ dex per dex).
Morover, within their M dwarf sample they found occurrence to {\it
  decrease} with mass by a factor of 2.7 relative to the prediction of
the universal relation of \citet{Johnson2010}, albeit at low
statistical significance.  \citet{Clanton2014} combined microlensing
and RV surveys to probe planet occurrence at similar distances and
found that M dwarfs have 2-3 times fewer giant planets than FGK
dwarfs.  However part of this difference is a result of their
assumption that comparisons should be scaled by the theoretical ``ice
line'' in the protoplanetary disk.  This is expected to lie at greater
distances from solar-type stars than M dwarfs and thus such a scaling
includes planets on a wider range of separations around the former.

\citet{Gaidos2013b} reported an occurrence of $4.0 \pm 2.3$\% for
Doppler-detected giant planets with orbital period $P < 245$~dy
orbiting late K dwarfs.  If extrapolated to the separation range
considered by \citet{Johnson2010} using a flat log-period
distribution, this corresponds to an occurrence of $5.5\pm3$\%.
\citet{Gaidos2013b} also performed a linear least-squares regression
of giant planet occurrence vs representative stellar mass from
published Doppler surveys.  The fit predicts a quadrupling of the
occurrence from M dwarfs to solar-mass stars, but the scatter is so
large that the significance of any mass-dependence is weak (F-test $p
= 0.12$).

One wrinkle in this story is the suggestion of \cite{Lloyd2011} that
the masses assigned to the evolved stars in the \citet{Johnson2010}
sample ($M_* > 1.5$\msun{}) have been overestimated and are only
slightly larger than solar.  The crux of the argument revolves around
whether there are too many high-mass stars in Doppler surveys
compared to plausible stellar population models
\citep{Lloyd2013,Johnson2013}.  \citet{Schlaufman2013} found that the
velocity dispersion of the stars in question were consistent with an
older age, and thus progenitors with a lower mass comparable to the
solar value.  This could mean that the trend with mass observed by
\citet{Johnson2010} is the result of a sensitive threshhold effect, an
artifact of age and orbital evolution, and/or the result of a high
false positive rate among evolved stars.

Another wrinkle is a systematic that arises when stars of different
types are compared: conflation of the effects of mass and metallicity.
The sensitivity of giant planet occurrence to metallicity means that
comparisons between stars of different masses must carefully control
for this parameter: an offset of only 0.1-0.2 dex in [Fe/H] could
equate to a factor of two in occurrence \citep{Neves2013}.  The
metallicity distribution of M dwarfs in Doppler surveys need not
necessarily be identical to that of solar-type stars in the solar
neighborhood because of differences in age and formation location or,
because metallicity and luminosity are related, due to selection bias.
The metallicities of M dwarfs have not been, until recently,
determined to the necessary precision, partly because the strength of
metallicity-sensitive features is difficult to determine at visible
wavelengths because of overlapping lines and the lack of a
well-defined continuum \citep[e.g.][]{Mann2013,Pineda2013}.  In
addition, calibrations based on stars with independently-established
metallicities did not exist.  This situation has been ameleriorated
with the proliferation of infrared spectrographs that probe a
wavelength range (1.2-2.5$\mu$m) of M dwarf spectra where there are
isolated metallicity-sensitive lines and a well-defined continuum
\citep{RojasAyala2012,Onehag2012,Terrien2012,Mann2013}.

To address the possible conflation of metallicity and mass dependence
we obtained infrared spectra of most of the M dwarfs monitored by the
California Planet Search (CPS).  We also obtained visible-wavelength
spectra of about half the sample to make new empirical estimates of
stellar parameters.  We describe the observations and data reduction
in Section \ref{sec.observations}, describe the resulting
distributions of metallicities and masses in Section
\ref{sec.results}, analyze and compare planet occurrence to the
well-characterized solar-mass stars in the Spectroscopic Properties of
Cool Stars \citep[SPOCS][]{Valenti2005} in Section \ref{sec.analysis},
and discuss our findings, their caveats, and their implications in
Section \ref{sec.discussion}.

\section{Observations and Reduction}
\label{sec.observations}

\subsection{SpeX Infrared Spectra}
\label{sec.spex}

Using the SpeX spectrograph \citep{Rayner:2003lr} on the NASA Infrared
Telescope (IRTF), we obtained simultaneous $JHK$ spectra of 121 of the
147 stars in the \citet[][hereafter RM06]{Rauscher2006} catalog of M
dwarfs monitored by the California Planet Search.  Observations were
carried out between September 2011 and December 2013 and targets were
selected entirely based on visibility from Maunakea, i.e. without
regard to their properties or whether they host known planets.  We
used SpeX in the cross-dispersed (SXD) mode and the 0.3~arcsec slit,
which yielded simultaneous coverage from 0.8-2.4$\mu$m at a resolution
$\lambda/\Delta \lambda$ of about 2000. Targets were nodded between
two positions along the slit (A and B) to ensure accurate sky
subtraction. At least six exposures were taken of each target
following the ABBA nodding pattern. Exposure times are capped at 120~s
to mitigate errors from atmospheric variation, although most exposure
times were much shorter (typically 10-20~s). This was sufficient to
ensure SNR of $>60$~pix$^{-1}$ in the $K$-band, and $>80$~pix$^{-1}$
in the $H$-band for all targets. We observed an A0V-type star, used to
remove telluric lines, within 1~hr of H.A. and 0.1~airmasses of each
target.

Reduction was carried out using the \texttt{SpeXtool} IDL package
\citep{Cushing:2004fk} which performed bias and flat-field
corrections, wavelength calibration, sky subtraction, and extraction
of the 1D spectrum. Multiple exposures were stacked using the
\texttt{xcombspec} routine. Telluric corrections were derived for each
target using the relevant A0V star observation and the
\texttt{xtellcor} package \citep{Vacca:2003qy}. Reduced spectra were
put in vacuum wavelengths and shifted to their rest wavelengths using
template stars from the IRTF spectral library \citep{Cushing:2005lr,
  Rayner:2009kx}.

Metallicities of the M dwarfs were calculated following the procedure
of \citet{Mann2013}.  They provide empirical relations between the
metallicity of M dwarfs and the strength of atomic lines in visible,
$J$, $H$, and $K$-band spectra, calibrated using wide binaries with an
F, G, or K dwarf primary and an M dwarf companion.  Metallicities of
the FGK primaries were forced to match the metallicity scale from
SPOCS \citep{Valenti2005} ensuring that the M dwarf metallicities are
on the same scale. We derived metallicities using the weighted mean of
the $H$ and $K$ band relations, accounting for both measurement and
errors in the calibration. Because the SNR of our SpeX data is
typically high ($>100$~pix$^{-1}$) errors are dominated by the
\citet{Mann2013} calibration errors, and the gain of adding in the
visible and $J$ band relations (which have higher errors) is
negligible. Resulting metallicities are reported in Table~1.

\subsection{SNIFS Visible Wavelength Spectra}
\label{sec.snifs}

We obtained visible wavelength spectra of 72 of the 147 stars in the
\citet{Rauscher2006} sample using the SuperNova Integral Field
Spectrograph \citep[SNIFS,][]{Aldering2002,Lantz:2004}
attached to the University of Hawaii 2.2m telescope atop
Maunakea. SNIFS utilizes separate blue (3200-5200~\AA) and red
(5100-9700~\AA) channels separated by a dichroic at resolutions of
$R\sim800$ and $R\sim1000$, respectively. Observations were done
between July 2010 and December 2012, primarily as part of
spectroscopic followup of the \citet{Lepine:2011vn} sample (Gaidos et
al. submitted). Exposure times ranged from 30-270s, which provided
peak SNR of $>120$~pix$^{-1}$ in the red channel for all
targets. Although the SNR was lower in the blue channel, these data
were not used for our analysis.

Basic SNIFS reduction, including bias and flat field corrections,
wavelength calibration, masking cosmic rays and bad pixels, and
extracting the 1-dimensional spectrum from the data cube, were
performed by the SNIFS reduction pipeline as described in
\citet{Bacon:2001} and \citet{Aldering:2006}. Spectra were then flux
calibrated using the atmosphere model from \citet{Buton:2013} combined
with spectrophotometric standards observed throughout the night
\citep{Oke1990, Hamuy1994,Bohlin2001}. Spectra were then shifted to
their rest frames by putting the wavelengths in vacuum, then
cross-correlating each spectrum to a template from
\citet{Bochanski:2007} of the corresponding spectral type. More
details on our reduction can be found in \citet{Lepine:2013}.

We determined the physical parameters of the stars with SNIFS spectra
using a modified version of the procedure described in
\citet{Boyajian2012,Mann2013c}.  This procedure performs a best-fit match between a
stellar spectrum and spectra generated by the PHOENIX stellar
atmosphere model \citep{Rajpurohit2014}.  A fixed set of wavelength
interval where the observations and model disagree are excluded in
such a way as to achieve agreement between best-fit values of
effective temperatures \teff{} and the bolometric temperatures of
calibrator stars established by measuring angular radii and bolometric
fluxes \citep{Boyajian2012,Mann2013c}.  We used the BT-SETTL grid of
model atmospheres based on the solar abundances of \citet{Caffau2010}.
We incorporated the [Fe/H] determined from our SpeX spectra (Section
\ref{sec.spex}) as a constraint on the fit by adding an additional
term to $\chi^2$.  We constructed average spectra from
randomly-selected sets of three of the best-fit grid points to
identify better fits in an interpolated grid.  Error in \teff{} was
determined by adding random noise to each spectrum according to the
formal error. We added 60~K error in quadrature to represent the
``floor'' in error from the calibraton itself.

For stars without visible-wavelength spectra, we estimated \teff{}
using spectral curvature indices calculated from the SpeX $K$-band
spectra and the calibration described in \citet{Mann2013c}.  Formal
errors in the indices were calculated using 100 Monte Carlo
realizations, however the error in \teff{} is dominated by the
residual 73K error in the calibration \citep{Mann2013c}.

We converted values of \teff{} into radii, luminosities, and masses
using the metallicity-independent empirical relations of
\citet{Mann2013c}.  Errors in these parameters were calculated based
on the error in the slope combined with the slope of each empirical
curve, plus the error in the calibrations added in quadrature.  The
coolest star in our calibration has \teff{} = 3238K; the other
parameters of RM06 stars with \teff{} below this value were assigned
upper limits.  Values are reported in Table~1.

\section{Metallicities and Masses of M dwarfs in the Solar Neighborhood}
\label{sec.results}

\begin{figure}
\includegraphics[width=84mm]{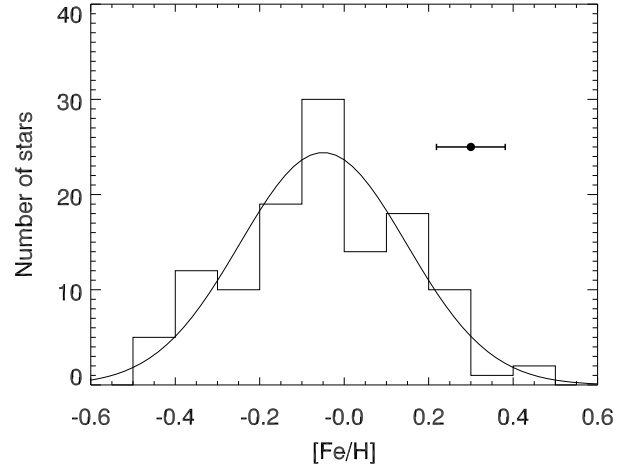}
\caption{Distribution of [Fe/H] for 121 M dwarf stars based on SpeX
  infrared spectra.  The curve is a Gaussian fit to the distribution
  with a mean of [Fe/H] = -0.05 and $\sigma$ of 0.20 dex.}
 \label{fig.fedist}
\end{figure}

The distribution of spectroscopic metallicities of 121 M dwarfs from
the RM06 catalog is plotted in Fig. \ref{fig.fedist}.  The mean and
median are $-0.050 \pm 0.008$ and $-0.060 \pm 0.014$, respectively,
with uncertainties determined by Monte Carlo simulation. The standard
deviation is 0.20 dex and the intrinsic deviation (after substracting
formal errors in quadrature) is 0.18 dex.  The distribution is
well-described by a Gaussian (Kolmogorov-Smirnov test probability 0.93
that the sample is drawn from the best-fit Gaussian).  All of these
values are very close to those established using visible-wavelength
spectra on a much larger sample of bright nearby M dwarfs (Gaidos et
al., submitted)

\begin{figure}
\includegraphics[width=84mm]{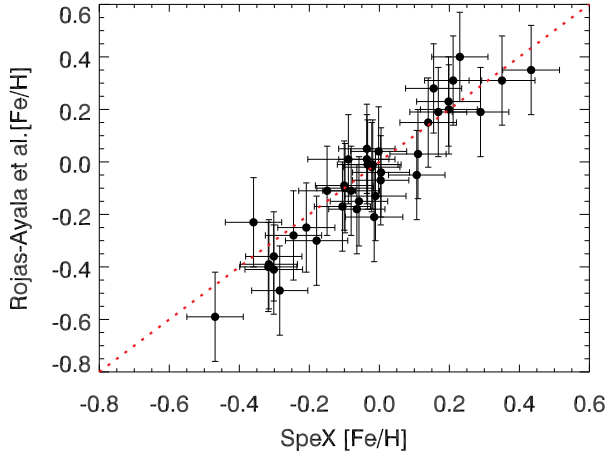}
\caption{Comparison of 39 M dwarf metallicities based on SpeX infrared
  spectra (this work) with those of \citet{RojasAyala2012}.  The
  dashed red line is equality.  The mean difference is $0.03 \pm 0.03$
  dex and the standard deviation is 0.09 dex.}
 \label{fig.rojas}
\end{figure}

There have been other studies of the metallicities of nearby M dwarfs
in exoplanet surveys, including
\citet{Johnson2009,Schlaufman2010,Rojas-Ayala2010}.  The most similar
study is that of \citet{RojasAyala2012}, which also used infrared
spectra and some of the same lines used here.  Figure \ref{fig.rojas}
shows excellent agreement between estimates for the 39 stars that
overlap with our RM06 sample.  The mean difference is statistically
insignficant ($0.03 \pm 0.03$) and the reduced chi-squared $\chi_{\nu}
^2$ of 0.23 suggests that our formal errors (and/or theirs) are too
conservative.

The largest studies have been those of \citet{Neves2012} and
\citet{Neves2013} of the CPS sample and southern M dwarfs observed by
the ESO/HARPS spectograph \citep{Bonfils2013}.  \citet{Neves2012} used
the spectroscopically-determined metallicities of primaries in solar
type-M dwarf binaries to refine the photometric M dwarf-metallicity
calibration that \citet{Schlaufman2010} developed based on $V$-$K_s$
colors and absolute $K_s$ magnitudes.  They then applied this to the
CPS sample \citep{Neves2013}.  They reported a residual dispersion
between the spectroscopic values of the solar-type primaries and the
photometric values of the M dwarf secondaries of 0.17~dex, which we
adopted as their formal error.  Separately, \citet{Neves2013}
correlated lines or features in HARPS visible-wavelength echelle
spectra with the photometric metallicities of a sample of M dwarfs.
They applied this relation to 102 M dwarfs in the HARPS Guaranteed
Time Observations sample \citep{Bonfils2013}.  They gave a calibration
dispersion of 0.08~dex and we adopted this as the formal error.

\begin{figure}
\includegraphics[width=84mm]{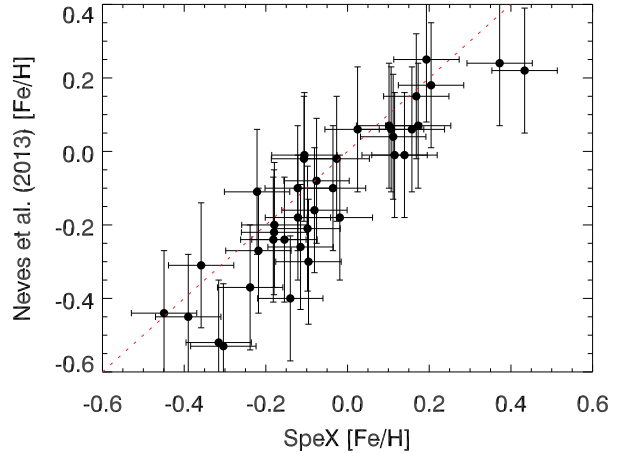}
\caption{Comparison of 115 M dwarf metallicities based on SpeX
  infrared spectra (this work) with those based on calibrated
  photometry \citep{Neves2012}.  The dashed red line is equality.
  The mean difference is $0.08 \pm 0.02$ dex and the standard
  deviation is 0.1 dex.}
 \label{fig.neves}
\end{figure}

\begin{figure}
\includegraphics[width=84mm]{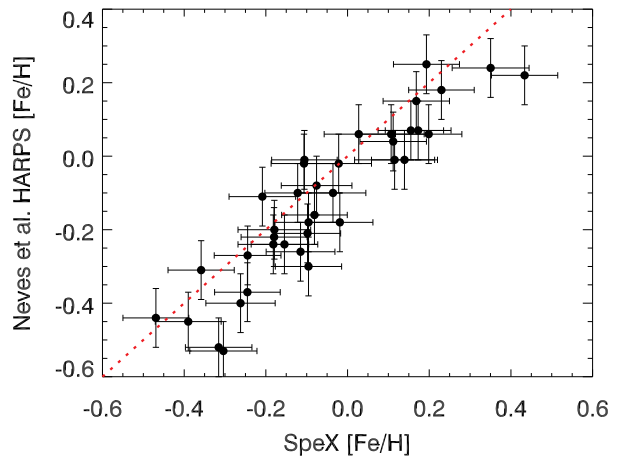}
\caption{Comparison of 38 M dwarf metallicities based on SpeX infrared
  spectra (this work) with those based on HARPS visible-wavelength
  spectra \citep{Neves2013}.  The dashed red line is equality.  The
  mean difference is $0.06\pm 0.02$ dex and the standard deviation is
  0.09 dex.}
 \label{fig.harps}
\end{figure}

We have metallicities for 115 and 38 stars in the
\citet{Neves2012} and \citet{Neves2013} samples.  Our values are
compared with theirs in Figs. \ref{fig.neves} and \ref{fig.harps}.
The weighted mean differences (this work - Neves) are $0.08 \pm 0.02$
and $0.06 \pm 0.02$, respectively.  The respective standard devations
are 0.10 and 0.08 dex, corresponding to $\chi_{\nu}^2$ values of 0.28
and 0.58.  The small offsets are consistent with the finding by
\citet{Sousa2008} that the method of \citet{Santos2004} to determine
the metallicities of the primary stars used by \citet{Neves2012} is on
the same scale as the SPOCS catalog, and that the spectroscopic scale
of \citet{Neves2013} is tied to the photometric calibration of
\citet{Neves2012}.  In addition, there is significant overlap in the
solar type-M dwarf binary calibrators used by \citet{Neves2013} and
\citet{Mann2013}.  The low $\chi^2_{\nu}$ values may indicate that
errors are overestimated. 

\begin{figure}
\includegraphics[width=84mm]{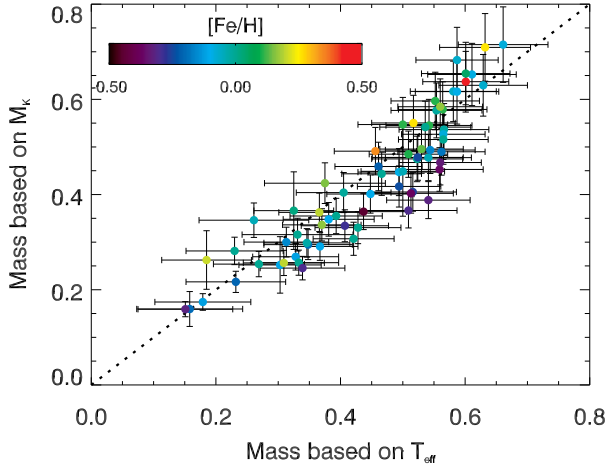}
\caption{Comparison of estimated mass of M dwarfs based on
  spectroscopic estimation of \teff{} vs. those based on absolute
  $K_S$-band magnitudes and the mass-luminosity relation of
  \citet{Delfosse2000}.  Points are colored according to metallicity.
  The dashed line is equality.  The mean of the ratio is 1.07 and the
  standard deviation is 0.15}
 \label{fig.mass_compare}
\end{figure}

Of the 121 stars in our sample, 118 have parallaxes.  We estimated
masses using the $M_K$-mass relation in \citet{Delfosse2000} and
compared these with values based on our \teff{} estimates (Section
\ref{sec.snifs}) and the \teff{}-mass relation in \citet{Mann2013c}.
    [That relation in turn is based on masses of calibrator M dwarfs
      from \citet{Boyajian2012} using the mass-luminosity relation of
      \citet{Henry1993}].  There is good agreement between the
    estimates (Fig. \ref{fig.mass_compare}), with a mean ratio of the
    \teff{}-based over the $M_K$-based estimates is 1.07, the standard
    deviation is 0.15, and $\chi_{\nu}^2$ is 0.86.  There is a
    possible trend with metallicity, with the temperatures (and thus
    masses) of metal-poor stars based on \teff{} estimates exceeding
    the estimates based on $M_K$ and \citet{Delfosse2000}.  This plot
    does not establish which method is effect by this systematic; if
    it exists, however, a plot of \teff{} from fits to PHOENIX models
    vs. values derived from the curvature of the $K$-band spectrum
    \citep{Mann2013c} show no trend with [Fe/H].  The $M_K$-based
    estimates have slightly smaller formal errors and we use them in
    our subsequent analysis.  For the three stars without parallaxes
    or visible-wavelength spectra we used the $K$-band-based masses.

\section{Analysis: Planets, Masses, and Metallicities}
\label{sec.analysis}

We compared metallicities and giant planet occurrence in the M dwarf
sample to that of the SPOCS sample \citep{Valenti2005}.  Critically,
and as discussed in Section \ref{sec.observations}, our M dwarf
metallicities are calibrated against solar-type primaries which have
been analyzed in an identical manner to SPOCS \citep{Mann2013} and for
which the offset of values for overlapping SPOCS and calibrator stars
is $< 0.01$ dex.  In fact, some of the calibrator primaries {\it are}
SPOCS stars.  Thus any systematic error in SPOCS metallicities is also
present in our M dwarf metallicities and to zero order is removed.
Because our M dwarf calibration was performed over a range of [Fe/H]
similar to that of SPOCS, it is removed to at least first order in
[Fe/H] as well.  The median and mean [Fe/H] of SPOCS stars are +0.04
and -0.01, respectively, and thus are, on average, slightly more
metal-rich than the statistics of our RM06 M dwarf sample.

\begin{figure}
\includegraphics[width=84mm]{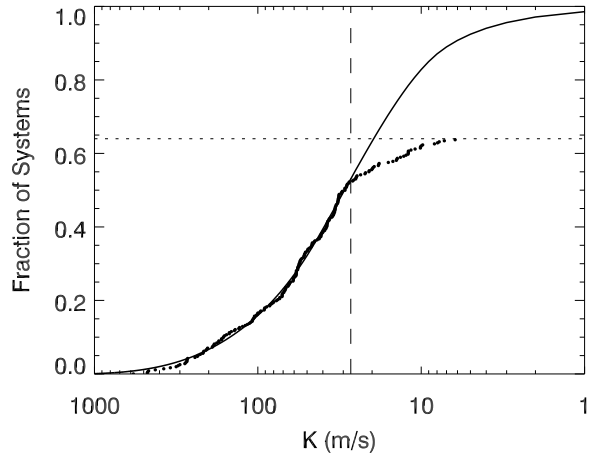}
\caption{Cumulative distribution of the RV amplitude $K$ of all
  Doppler-detected giant planets with $0.1M_J < M \sin i < 10M_J$ and
  $P< 2$yr in the Exoplanet Orbit database \citep{Wright2011}. The
  vertical dashed line at $K = 37$~\mps{} marks a break in the
  distribution we interpret to indicate the onset of detection
  incompleteness.  The solid curve is a best-fit for a planets with
  power-law mass-distribution index of 0.78 over $0.3-10M_J$, a flat
  distribution with log orbital period over $2 {\rm d} < P < 2 {\rm
    yr}$, a Rayleigh-distributed eccentricity distribution with mean
  $\bar{e} = 0.25$, and isotropically inclined orbits.}
 \label{fig.complete}
\end{figure}

We obtained current planet data from the Exoplanet Orbit Database
\citep{Wright2011}.  For consistency with previous analyses we
consider giant planets with projected masses $0.1M_J < M \sin i <
10M_J$ and orbital periods $P < 2$~yr.  The lower mass limit removes
Neptune-size planets which are considered a distinct population of
objects with a difference formation pathway \citep{Benz2014}.  We
estimated the completeness of the exoplanet archive catalog with RV
amplitude $K$ from the cumulative distribution of detected planets
itself (Fig. \ref{fig.complete}).  We identified the break in the
distribution at $K \approx 27$ \mps{} as the point below which the
catalog becomes incomplete for the planet mass and period range of
interest.  A planet with $M \sin i = 1M_J$ on a 2~yr circular orbit
around a solar-mass star produces $K \approx 27$~\mps{}, a signal
which can be readily detected in high-precision Doppler surveys,
provided there is sufficient time baseline.  To calculate the
completeness of the catalog over some actual mass range, accounting
for the underlying distribution of eccentricities and orbital
inclinations, we assume that the distribution with $M$ and $P$ are
smooth functions of the usual form:
\begin{equation}
\label{eqn.massperiod}
dN = M^{-\alpha}P^{-\beta} d\log M\,d\log P
\end{equation}
We assumed a Rayleigh distribution for orbital eccentricities with a
mean $\bar{e}$ \citep{Moorhead2011a} and isotropically-distributed
orbital inclinations.  We determined the completeness $C$ of the
catalog by calculating the cumulative distribution of $K$ for a
population of Saturn- to super-Jupiter-masses ($0.3M_J < M < 10M_J$)
and periods $3{\rm dy} < P < 2$yr, and finding best-fit values of
$\alpha$, $\beta$, $\bar{e}$, and $C$ by non-linear least-squares
fitting with the MPFIT routine \citep{Markwardt2009}.  We find $\alpha
= 0.78$, $\beta = -0.003$, $\bar{e} = 0.25$ consistent with previous
findings \citep{Cumming2008}, and $C = 0.64$.  Thus our adopted
subsample of the catalog captures the majority of Saturn-mass and
larger objects to 2~yr periods.  We use a single value of $C$ to
describe our completeness.

There are 73 giant planets satisfying our criteria among 1039 SPOCS
stars.  We identified giant planets orbiting three RM06 stars for
which we have determined [Fe/H]: GJ~1148, GJ~649, and GJ~876, with
metallicities of $-0.04\pm0.08$, $0.00\pm0.08$, and $+0.17\pm0.08$,
respectively.  \citet{RojasAyala2012} found metallicities of
$+0.05\pm0.17$, $-0.04\pm0.17$, and $+0.19\pm0.17$, respectively, in
excellent agreement with our values.  Likewise, \citet{Neves2013}
found metallicities of +0.07, -0.08, and +0.12.

\begin{figure}
\includegraphics[width=84mm]{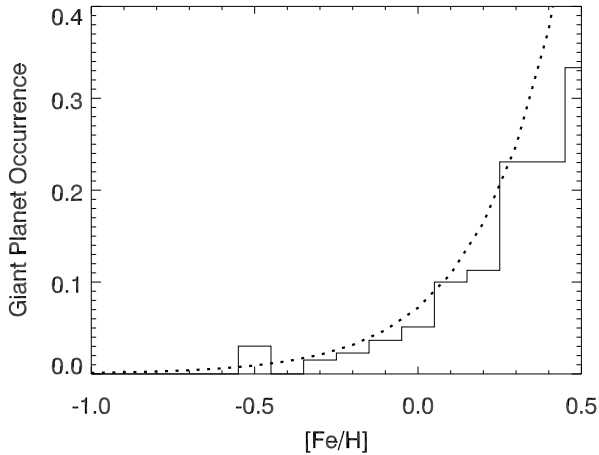}
\caption{Occurrence of giant planets ($0.1M_J < M \sin i < 10M_J$, $P<
  2$yr) among SPOCS stars vs. metallicity.  The dashed line is
  Eqn. \ref{eqn.planetmetal} with maximum likelihood coefficients 1.80
  dex per dex and an occurence of 0.070 at solar metallicity, and a
  completeness of 0.651.}
 \label{fig.spocs}
\end{figure}

We infer a planet-metallicity relation of the standard form
\begin{equation}
\label{eqn.planetmetal}
f({\rm [Fe/H]}) = f_0 10^{a{\rm [Fe/H]}}
\end{equation}
and determine the values of $f_0$ and $a$ that maximize the binomial likelihood, i.e.
\begin{equation}
\label{eqn.likelihood}
\log \mathcal{L} = \sum_i^{D} \log f({\rm [Fe/H]_i}) + \sum_j^{ND} \log \left[1-C \, f({\rm [Fe/H]_j})\right],
\end{equation}
where the first and second summations are over $D$ and $ND$ systems
with and without detected giant planets, respectively.  (The factor
$C$ in the first summed logarithms contributes only a constant term to
the likelihood and is thus omitted.)  For the SPOCS stars we found
$f_0 = 0.070$ and $a = 1.80$, consistent with \citet{Fischer2005}. Our
value of $f_0$ is slightly lower than \citet{Cumming2008} when
extrapolated to $P=2000$~d.  Fig. \ref{fig.spocs} shows that
Eqn. \ref{eqn.planetmetal} with these parameter values describes the
planet-metallicity distribution well.  The notable exception is
HIP~37124, which has [Fe/H] = -0.44 and three close-in giant planets
with a possible 2:1 mean-motion resonance \citep{Wright2011b}.  The
metal-poor nature of this star has been confirmed by independent
observations \citep{Santos2003,Kang2011}

We test the null hypothesis, i.e. that the planet-metallicity relation
derived for the SPOCS sample adequately describes giant planet
occurrence among M dwarfs.  The expected number of giant planets using
our values of [Fe/H] and Eqn. \ref{eqn.planetmetal} is 6.4.  Assuming
Poisson statistics, the probabilities of finding 3 or fewer planets is
0.12.  Since the equivalent of one-sigma probability is one half of
1-0.68, or 0.16 we conclude that the occurrence of giant planets
around M dwarfs differs from that of solar-type stars at slightly more
than $1\sigma$ significance.  Without a correction for the metallicity
distributions of the SPOCS and M dwarf samples we would expect 8.5
planets and the difference would be much more significant.

Although dependence on stellar mass is not required to explain the
observations, what mass-dependence can be {\it excluded}?  We assumed
the form
\begin{equation}
\label{mass.metal}
f = f_0 10^{a {\rm [Fe/H]}}M_*^b,
\end{equation}
where $M_*$ is in solar units and we consider $b \in [-1,2]$.  For
SPOCS stars we used the mass estimates from \citet{Valenti2005}.  As
discussed above, SPOCS mass assignments are controversial for
higher-mass, evolved stars \citep{Lloyd2013} and they are probably not
reliable for late K and M dwarf stars.  Therefore we only considered
the 815 stars (55 planets) with 0.7\msun{}$ < M_* < 1.5$\msun{} and
$\log g > 3.85$.

\begin{figure}
\includegraphics[width=84mm]{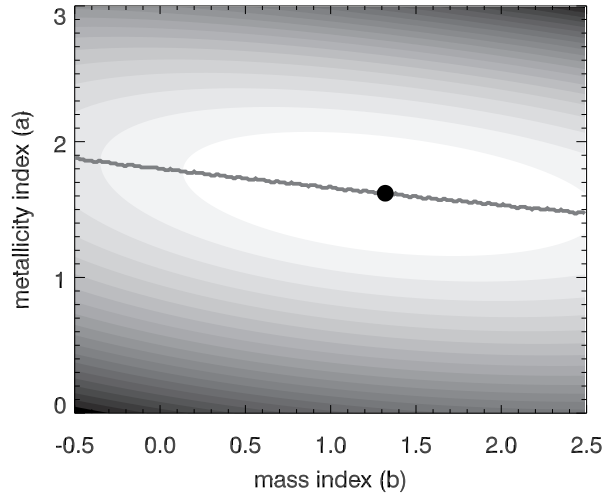}
\caption{Contour plot of $\log$ likelihood vs. metallicity and mass
  occurrence parameters $a$ and $b$ among SPOCS solar-type stars.
  Each shade represents one relative unit of $\log$ likelihood, with
  lighter shades representing higher values.  The black dot is the
  location of the maximum ($a = 1.62$, $b=1.32$) and the grey line is
  a ``ridge'' of values of $a$ that maximize the likelihood for a
  given $b$.}
 \label{fig.alphabeta}
\end{figure}

\begin{figure}
\includegraphics[width=84mm]{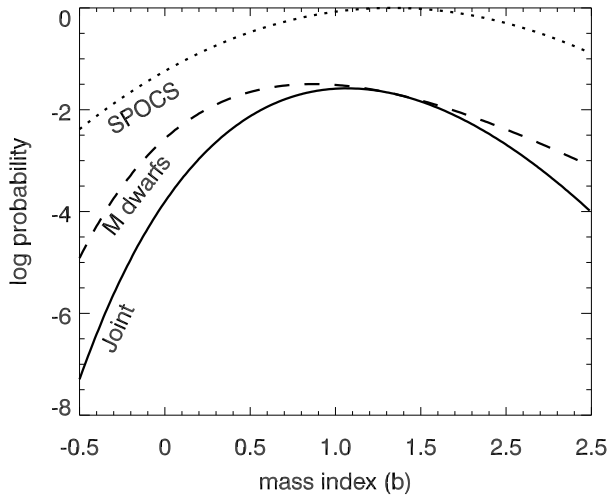}
\caption{Log likelihood relative to the maximum value for SPOCS stars
  (dashed line), logarithm of the probability of the observed number
  of giant planets in the RM06 M dwarf sample (dotted line), and the
  joint likehoods (solid line) vs. mass index $b$ and metallicity
  index $a$ along the trajectory in Fig. \ref{fig.alphabeta}.}
 \label{fig.beta}
\end{figure}

Figure \ref{fig.alphabeta} is a contour of the relative $\log$
likelihood vs. $a$ and $b$ for SPOCS stars.  Regions of parameter
space with high or low values of $a$ are definitively excluded, but a
wide range of $b$ is allowed.  The maximum is at $a \approx 1.62$ and
$b \approx 1.32$ (black dot), in agreement with \citet{Montet2014}.
The grey line is the ``ridge'' of values of $a$ maximizing likelihood
for a given value of $b$.  Fig. \ref{fig.beta} plots the relative
$\log$ likelihood along the ``ridge'' (dotted line).  Also plotted is
the probability of observing three planets in our M dwarf sample along
this trajectory in $a$-$b$ parameter space (dashed line).  This is
maximized at $b = 0.87$, consistent with the finding of
\citet{Montet2014}.  The solid line is the joint likelihood, assuming
that a single mass-metallicity-planet relation applies to all stars,
and it peaks at $b = 1.06$.  Assuming asympotic normality (AN,
parabolic behavior of the logarithmic likelihood near its maximum) for
the joint likelihood curve, the uncertainty in $b$ is $\pm 0.47$.  Thus
the data support a roughly linear stellar mass dependence but with
weak significance and any scenario between no mass-dependence ($b \sim
0$) and strong mass-dependence ($b \sim 2$) cannot be excluded.  If we
include the stars in the SPOCS sample with $M_* > 1.5$\msun{} and
accept those mass estimates, we find $b = 0.37 \pm 0.32$ and
significant mass-dependence is excluded because there are too few
planets around the most massive stars.  Clearly, reliable mass
estimates for these stars would be very valuable.

Finally, we independently evaluated the giant planet-metallicity
relation of M dwarfs and assess whether it is statistically
distinguishable from that of SPOCS solar-type stars.  An erroneous
assumption of a universal relation could mask mass dependence.  For
example, if the metallicity relation for M dwarfs were less steep than
for solar-type stars, a greater mass dependence would be required to
explain the relative deficit of giant planets.  We estimated
uncertainties using the assumption of AN for the likelihood curve as
well as by constructing 100 Monte Carlo representations of the data
assuming normally-distributed errors in metallicity; for the SPOCS
stars we assume an error of 0.03 dex for all stars
\citep{Valenti2005}.  For SPOCS stars we found $a = 1.80$ with an
uncertainty of $\pm 0.05$ based on Monte Carlo simulations, but 0.31
based on AN.  For the M dwarfs we found $a = 1.06$ with
uncertatainties of $0.42$ based on Monte Carlo and $\pm 1.25$ based on
AN.  (The best-fit $f_0$ is 0.025.)  The relative magnitudes of the
uncertainties indicates that it is sample size, rather than
metallicity precision, which limits statistical significance.  The
difference between values of $a$ for SPOCS stars and the M dwarfs is
less than 1$\sigma$ and not significant (see Section
\ref{sec.discussion}).

\section{Discussion and Conclusions}
\label{sec.discussion}

 We obtained infrared and visible-wavelength spectra to estimate the
 metallicities and other fundamental parameters of nearby M dwarfs
 monitored by the CPS Doppler exoplanet survey.  These M dwarfs are on
 average slightly more metal poor (0.1 dex) than the SPOCS comparison
 catalog of solar-mass stars that has also been monitored for planets.
 We then determined the number of Doppler-detected detected giant
 planets ($M \sin i = 0.1-10 M_J$) with orbital periods $<2$~yr around
 the M dwarfs as well as the solar-type stars.

{\it Heavy, or Metal?}  When we assumed that giant planet occurrence
depends only on metallicity, we derived a planet-metallicity relation
for solar-type stars (exponent $a = 1.80$ dex per dex) that
overpredicts occurrence around M dwarfs, but the discrepancy is not
statistically significant.  On the other hand, when we allowed both
mass and metallicity dependence, we found mass-dependence of $b \sim
1$ dex per dex for both solar-mass and M dwarf stars, but this is only
marginally significant and the hypothesis of no mass dependence cannot
be excluded.  We find a metallicity dependence for M dwarfs that is
lower than that of solar-mass stars ($a \approx 1$ rather than $a
\approx 1.8$) but this difference is, yet again, not significant due
to the small size of the M dwarf sample.

Because our M dwarf metallicities are calibrated against the
metallicities of primaries determined using the same technique as the
SPOCS catalog (and in fact, some of the calibrators are in the SPOCS
catalog), comparisons between the two samples are immune to any
systematic error in the masses and metallicities of SPOCS stars.
Nevertheless, the known covariance between estimates of [Fe/H] and
\teff{} (and hence mass) from SME \citep{Torres2012} still effects our
estimates of metallicity and mass indices $a$ and $b$.  Independent
estimation of the \teff{} of SPOCS stars using the spectroscopic,
model-independent methods described in \citet{Mann2013c}, which in
turn are based on the calibration stars studied by
\citet{Boyajian2012}, and then revision of [Fe/H] based on those
\teff{} values are warranted.

In contrast with \citet{Johnson2010}, only one of the three giant
planet hosts in our M dwarf sample is metal-rich with respect to the
sample mean.  Our values agree with previous estimates for these stars
based on infrared spectra \citep{Rojas-Ayala2012} or calibrated
photometry \citep{Neves2013}.  Curiously, if we relax our orbital
period restriction, two more giant planet hosts enter our sample:
GJ~179b (6.26~yr orbit) and GJ~849b (5.16~yr orbit).  Both of these
are metal-rich, but two objects do not make a trend.

Our M dwarf planet-metallicity relation is more shallow ($a = 1.05$)
then found by others \citep[$a \approx
  2-3$][]{Johnson2010,Neves2013,Montet2014}.  The discrepancy with
\citet{Neves2013} is particularly puzzling since there is excellent
agreement between the two sets of metallicity estimates
(Fig. \ref{fig.neves}).  Some part of the difference could be
explained by the 0.08 dex offset in mean [Fe/H], but also the exact
distribution of metallicities may matter.  As an experiment, we
substituted 115 values for [Fe/H] from \citet{Neves2013} into our
analysis and found $a = 1.99$, almost exactly the value of 1.97 found
by \citet{Neves2013}.  Thus the discrepancy is not due to any
methodological difference but instead arises from the sensitivity of
maximimum-likehood estimates of $a$ to the exact distribution with
[Fe/H].  This is also suggested by the large AN-based uncertainties in
$a$.  Support for a lower value of $a$ also comes from microlensing
surveys of the metal-rich Galactic Bulge, which find an order of
magnitude fewer giant planets around M dwarfs than predicted by a
steep planet-metallicity dependence \citep{Clanton2014}.

To explore the effects of sample size and measurement error, we
carried out Monte Carlo simulations of the recovery of $a$ from
synthetic populations of stars and planets.  Each simulated a sample
of 121 stars to which random Gaussian-distributed metallicities were
assigned.  Planets were placed around stars with a probability given
by Eqn. \ref{eqn.planetmetal} and fixed value of $a = 2$.  Then
Gaussian-distributed measurement errors were added to the actual
metallicities, and maximum-likelihood values of $\alpha$ were
recovered using Eqn. \ref{eqn.likelihood}.  In the absence of
measurement error, the mean of the inferred values of $a$ was close to
the actual value, but the standard deviation was $\sigma \approx 1$.
The large value of $\sigma$ agrees with our AN-based estimates of the
uncertainty in $a$.  The introduction of measurement errors lowers the
inferred value of $a$ relative to the actual value.  For example, if
the observed dispersion of 0.20 dex is the product of an intrinsic
variation of 0.18 dex and measurement error equal to 0.08 dex then the
mean value inferred from $\alpha = 2$ simulations is $\alpha = 1.85$.
The larger the contribution of error to the dispersion, the greater
the difference.  This effect alone, however, cannot explain the
difference between our results and \citet{Neves2013} because our
errors are similar in magnitude.

At orbital periods $\gg 2$~yr Doppler surveys become very incomplete,
i.e. missing the lower-mass ($\lesssim 1M_J$) majority of the giant
planet population.  This can be partially ameliorated if long-term
trends are included \citep{Montet2014}.  On the other hand,
microlensing surveys can already detect such planets, albeit around a
population of distant M dwarfs in the direction of the Galactic Bulge.
\citet{Clanton2014} performed a joint analysis of radial velocity and
microlensing surveys and estimated that $15 \pm 6$\% of M dwarfs host
planets with $M \sin i > 0.1M_J$ and $P < 27$~yr.  However, two-thirds
of these orbit outside 2.7~yr, demanding a marked departure from a
flat $\log P$ distribution.

{\it Kepler Weighs In:} The NASA \kep{} mission could readily detect
giant planets around both solar-type stars and M dwarfs, but the
former lie at kpc distances and their metallicities may not follow the
distribution in the solar neighborhood.  Moreover, while the [Fe/H] of
some individual solar-type host stars has been determined, the
underlying metallicity distribution has not yet been established.  In
contrast, \kep{} M dwarfs are a factor of 10 closer and are likely to
resemble their nearby counterparts \citep{Gaidos2012}.
\citet{Mann2013b} found a median [Fe/H] of $-0.10 \pm 0.03$ among non
KOI hosts, only slightly more metal poor than the RM06 sample.  We
calculated the expected occurrence of giant planets around \kep{} M
dwarfs based on the SPOCS distribution and find it to be 0.071.
Adjusting to $2 < P < 365$~d assuming a flat distribution with $\log
P$, the predicted occurrence is 0.06.

We determined the observed occurrence of giant planets ($R_p >$
6\rearth{}) around 3828 \kep{} stars with 2800K $<T_{\rm eff}<$ 4200K
and $\log g > 4.2$, stellar parameter that were estimated by Bayesian
analysis of photometry in the \kep{} Input Catalog \citep{Brown2011a},
the Dartmouth stellar evolution models \citep{Dotter2008}, and priors
on stellar, metallicity, age, mass, and distance \citep{Gaidos2013a}.
We restricted the sample to orbits with period $P<1$~yr and assumed,
based on the large transit depths ($> 1$\%), that all giant planets
which transit would be detected.  We assumed Rayleigh-distributed
eccentricities with a mean of 0.272 (see above) and calculated the
mean transit probability $p$ for each star assuming a flat logarithmic
distribution with $P$.  We then found the occurrence that maximized
the binomial log likelihood
\begin{equation}
\ln \ell = N_D \ln f + \sum_i^{ND} \left(1 - f p_i\right), 
\end{equation}
where $N_D$ is the number of detections, and the sum is over all stars
without detections.  We identify two candidate giant planets and find
that the likelihood is maximized at $f = 0.036$ with an uncertainty of
$\pm 0.021$ based on AN.  This is additional tantalizing but not
significant ($1.1\sigma$) evidence for a deficit of giant planets
around M dwarfs relative to solar-type stars.

{\it Is Theory in Irons?} Naively, if the formation of giant planets
depends on the mass of solids in a protoplanetary disk, and this
scales with both stellar mass and metallicity, then one would expect
that the planet-metallicity relation for M dwarfs would be {\it
  steeper} than that of solar-mass stars.  While our analysis cannot
exclude similar values of $a$ for M dwarfs and SPOCS stars, they seem
to exclude $a \gtrsim 3$, as does the microlensing analysis of
\citet{Clanton2014}.  Perhaps it is not the mass of solids but some
other factor, such as disk lifetime, that regulates giant planet
formation \citep{Yasui2009,Owen2012}.  This is certainly suggested by
the apparent absence of a planet-metallicity relation for Neptune-size
planets that could serve as the cores of giant planets
\citep{Neves2013,Mann2013b}.  

This analysis, and essentially all others, is predicated on simple and
universal relationsips between planet occurrence and stellar
mass/metallicity.  Nature may not be so compliant.  Orbital migration
is now widely accepted to play an important role in determining the
distribution of giant planets, particularly those close to their host
stars and hence over-represented in Doppler and transit surveys.
However, many aspects of that migration are not well understood,
including the particularly salient aspect of what halts it
\citep{Hasegawa2013}.  Analysis of microlensing surveys suggest that
the majority of giant planets orbiting M dwarfs are found beyond $\sim
1.5$~AU and these must be considered, and hopefully included
\citet{Montet2014} when elucidating the planet-metallicity
relationship of such stars.

In our analysis we have assumed that the mass distribution
(Eqn. \ref{eqn.massperiod}) of giant planets orbiting M dwarfs is the
same as that around solar-type stars.  If M dwarfs have less massive
disks, or the gas dissipates more rapidly, giant planet formation
would terminate at lower masses, producing a steeper mass distribution
with a larger value of $\alpha$, a possibility suggested by
\citet{Neves2013}.  Doppler surveys, which are biased with mass, would
then find an apparent deficit of giant planets around M dwarfs
relative to solar-type stars, even if the actual occurrence is the
same.

{\it Massive Improvements Ahead:} Our analysis, along with every
previous one, is limited by the small comparatively small number of M
dwarfs in exoplanet surveys.  A modest but important step will be
spectroscopic measurement of [Fe/H] in all M dwarf stars in Doppler
radial velocity surveys, especially the HARPS GTO survey
\citep{Bonfils2013}.  However, this will not even double the size of
the sample described here.  The advent of high-precision spectrographs
operating at infrared wavelengths where there is considerably more
signal from M dwarfs will usher in larger Doppler radial velocity
surveys \citep{Tamura2012,Thibault2012,Quirrenbach2012,Mahadevan2012}.  These
surveys are typically envisioned to include $\sim$300 stars and proper
coordination could lead to an order-of-magnitude improvement in sample
size.  An important consideration is the time baseline needed to
detect planets over the orbital period range considered here, well
beyond the M dwarf ``habitable zone'' whose compact size drives the
design of these surveys.  An alternative strategy is to survey the
most metal-rich M dwarfs with a Dopler survey, thus taking advantage
of a longer ``lever arm'' with which to constrain the mass-metallicity
index $a$.  We are pursuing such a survey at visible and, in the
future, infrared wavelengths.

The \kep{}-2 Mission \citep{Howell2014} and NASA Transiting Exoplanet
Survey Satellite mission \citep{Ricker2014} will make modest
contributions because of the short observation periods $\ll 1$~yr and
the strong bias of the transit method towards short-period orbits
where giant planets are rare around M dwarfs.  In contrast, the {\it
  Gaia} spacecraft will monitor thousands of nearby M dwarfs with an
astrometric astrometric precision of tens of $\mu$arcsec
\citep{deBruijne2012}.  \citet{Sozzetti2014} estimated that $\sim$100
giant planets would be detected around the nearest M dwarfs ($d
\lesssim 50$~pc).  Conveniently, {\it Gaia} will also provides precise
parallaxes with which $M_K$ and hence mass can be estimated.
Follow-up spectroscopy of M dwarfs with {\it Gaia}-detected companions
and/or the calibration of {\it Gaia} spectra of the Ca II infrared
triplet to determine metallicity should be a high priority.

\acknowledgments

This research was supported by NASA grants NNX10AQ36G and NNX11AC33G
to EG. It has made use of the Exoplanet Orbit Database and the
Exoplanet Data Explorer at exoplanets.org

\begin{table*}
\centering
\begin{minipage}{110mm}
\caption{Parameters of Nearby M Dwarfs from the California Planet Search \label{tab.params}}
\begin{tabular}{@{}lrrrrrrl@{}}
\multicolumn{1}{c}{Name} & \multicolumn{1}{c}{HIP} & \multicolumn{1}{c}{[Fe/H]} & \multicolumn{1}{c}{$T_{\rm eff}$ (K)} & \multicolumn{1}{c}{$R_*$ ($R_{\odot}$)} & \multicolumn{1}{c}{$L_*$ ($L_{\odot}$)} & \multicolumn{1}{c}{$M_*$ ($M_{\odot}$)} & \multicolumn{1}{c}{$T_{\rm eff}$ src} \\
 & & & & \multicolumn{3}{c}{(negative values are upper limits)} & \\
\hline
          GJ 2 &    428 & -0.01$\pm$0.08 & 3781$\pm$ 73 &  0.52$\pm$0.04 &  0.0482$\pm$0.0091 &  0.56$\pm$0.07 &  K-band \\
     HD 225213 &    439 & -0.39$\pm$0.08 & 3401$\pm$ 73 &  0.32$\pm$0.06 &  0.0131$\pm$0.0051 &  0.31$\pm$0.07 &  K-band \\
      LHS 1053 &   1368 &  0.07$\pm$0.08 & 4336$\pm$ 73 &  0.67$\pm$0.04 &  0.1462$\pm$0.0191 &  0.69$\pm$0.07 &  K-band \\
        GX And &   1475 & -0.26$\pm$0.08 & 3693$\pm$ 91 &  0.49$\pm$0.05 &  0.0382$\pm$0.0099 &  0.52$\pm$0.07 & Visible \\
        GQ And &      0 & -0.20$\pm$0.08 & 3254$\pm$ 71 &  0.20$\pm$0.07 &  0.0041$\pm$0.0039 &  0.16$\pm$0.09 & Visible \\
     BPM 46239 &   1734 &  0.27$\pm$0.08 & 3695$\pm$ 82 &  0.49$\pm$0.05 &  0.0384$\pm$0.0090 &  0.52$\pm$0.07 & Visible \\
       LHS 119 &      0 & -0.11$\pm$0.08 & 3523$\pm$ 73 &  0.40$\pm$0.05 &  0.0223$\pm$0.0062 &  0.41$\pm$0.07 &  K-band \\
      LHS 1122 &   3143 &  0.01$\pm$0.09 & 3596$\pm$ 73 &  0.44$\pm$0.05 &  0.0287$\pm$0.0069 &  0.46$\pm$0.06 &  K-band \\
     BD+61 195 &   4872 &  0.15$\pm$0.08 & 3835$\pm$ 73 &  0.54$\pm$0.04 &  0.0549$\pm$0.0099 &  0.58$\pm$0.07 &  K-band \\
        YZ Cet &   5643 & -0.26$\pm$0.08 & 3125$\pm$ 61 & -0.19$\pm$0.00 & -0.0033$\pm$0.0000 & -0.14$\pm$0.00 & Visible \\
\hline
\end{tabular}
\medskip
Table 1 is published in its entirety in the electronic edition of \emph{The Astrophysical Journal}. A portion is shown here for guidance regarding its form and content.
\end{minipage}
\end{table*}


\begin{thebibliography}{75}
\expandafter\ifx\csname natexlab\endcsname\relax\def\natexlab#1{#1}\fi

\bibitem[{{Aldering} {et~al.}(2002){Aldering}, {Adam}, {Antilogus}, {Astier},
  {Bacon}, {Bongard}, {Bonnaud}, {Copin}, {Hardin}, {Henault}, {Howell},
  {Lemonnier}, {Levy}, {Loken}, {Nugent}, {Pain}, {Pecontal}, {Pecontal},
  {Perlmutter}, {Quimby}, {Schahmaneche}, {Smadja}, \&
  {Wood-Vasey}}]{Aldering2002}
{Aldering}, G., {et~al.} 2002, in Proc. SPIE, ed. {J.~A.~Tyson \& S.~Wolff},
  Vol. 4836, 61--72

\bibitem[{{Aldering} {et~al.}(2006){Aldering}, {Antilogus}, {Bailey}, {Baltay},
  {Bauer}, {Blanc}, {Bongard}, {Copin}, {Gangler}, {Gilles}, {Kessler},
  {Kocevski}, {Lee}, {Loken}, {Nugent}, {Pain}, {P{\'e}contal}, {Pereira},
  {Perlmutter}, {Rabinowitz}, {Rigaudier}, {Scalzo}, {Smadja}, {Thomas},
  {Wang}, {Weaver}, \& {Nearby Supernova Factory}}]{Aldering:2006}
{Aldering}, G., {et~al.} 2006, \apj, 650, 510

\bibitem[{{Andrews} {et~al.}(2013){Andrews}, {Rosenfeld}, {Kraus}, \&
  {Wilner}}]{Andrews2013}
{Andrews}, S.~M., {Rosenfeld}, K.~A., {Kraus}, A.~L., \& {Wilner}, D.~J. 2013,
  \apj, 771, 129

\bibitem[{{Bacon} {et~al.}(2001){Bacon}, {Copin}, {Monnet}, {Miller},
  {Allington-Smith}, {Bureau}, {Carollo}, {Davies}, {Emsellem}, {Kuntschner},
  {Peletier}, {Verolme}, \& {de Zeeuw}}]{Bacon:2001}
{Bacon}, R., {et~al.} 2001, \mnras, 326, 23

\bibitem[{{Benz} {et~al.}(2014){Benz}, {Ida}, {Alibert}, {Lin}, \&
  {Mordasini}}]{Benz2014}
{Benz}, W., {Ida}, S., {Alibert}, Y., {Lin}, D.~N.~C., \& {Mordasini}, C. 2014,
  ArXiv e-prints

\bibitem[{{Bochanski} {et~al.}(2007){Bochanski}, {West}, {Hawley}, \&
  {Covey}}]{Bochanski:2007}
{Bochanski}, J.~J., {West}, A.~A., {Hawley}, S.~L., \& {Covey}, K.~R. 2007,
  \aj, 133, 531

\bibitem[{{Bohlin} {et~al.}(2001){Bohlin}, {Dickinson}, \&
  {Calzetti}}]{Bohlin2001}
{Bohlin}, R.~C., {Dickinson}, M.~E., \& {Calzetti}, D. 2001, \aj, 122, 2118

\bibitem[{{Bonfils} {et~al.}(2013){Bonfils}, {Delfosse}, {Udry}, {Forveille},
  {Mayor}, {Perrier}, {Bouchy}, {Gillon}, {Lovis}, {Pepe}, {Queloz}, {Santos},
  {S{\'e}gransan}, \& {Bertaux}}]{Bonfils2013}
{Bonfils}, X., {et~al.} 2013, \aap, 549, A109

\bibitem[{{Boyajian} {et~al.}(2012){Boyajian}, {von Braun}, {van Belle},
  {McAlister}, {ten Brummelaar}, {Kane}, {Muirhead}, {Jones}, {White},
  {Schaefer}, {Ciardi}, {Henry}, {L{\'o}pez-Morales}, {Ridgway}, {Gies}, {Jao},
  {Rojas-Ayala}, {Parks}, {Sturmann}, {Sturmann}, {Turner}, {Farrington},
  {Goldfinger}, \& {Berger}}]{Boyajian2012}
{Boyajian}, T.~S., {et~al.} 2012, \apj, 757, 112

\bibitem[{Brown {et~al.}(2011)Brown, Latham, Everett, \& Esquerdo}]{Brown2011a}
Brown, T.~M., Latham, D. W.~D., Everett, M. E.~M., \& Esquerdo, G. G.~A. 2011,
  Astron. J., 142, 112

\bibitem[{{Buton} {et~al.}(2013){Buton}, {Copin}, {Aldering}, {Antilogus},
  {Aragon}, {Bailey}, {Baltay}, {Bongard}, {Canto}, {Cellier-Holzem},
  {Childress}, {Chotard}, {Fakhouri}, {Gangler}, {Guy}, {Hsiao}, {Kerschhaggl},
  {Kowalski}, {Loken}, {Nugent}, {Paech}, {Pain}, {P{\'e}contal}, {Pereira},
  {Perlmutter}, {Rabinowitz}, {Rigault}, {Runge}, {Scalzo}, {Smadja}, {Tao},
  {Thomas}, {Weaver}, {Wu}, \& {Nearby SuperNova Factory}}]{Buton:2013}
{Buton}, C., {et~al.} 2013, \aap, 549, A8

\bibitem[{Caffau \& Freytag(2010)}]{Caffau2010}
Caffau, E., \& Freytag, B. 2010, Solar Physics, 1

\bibitem[{{Clanton} \& {Gaudi}(2014)}]{Clanton2014}
{Clanton}, C., \& {Gaudi}, S. 2014, ArXiv e-prints

\bibitem[{{Cumming} {et~al.}(2008){Cumming}, {Butler}, {Marcy}, {Vogt},
  {Wright}, \& {Fischer}}]{Cumming2008}
{Cumming}, A., {Butler}, R.~P., {Marcy}, G.~W., {Vogt}, S.~S., {Wright}, J.~T.,
  \& {Fischer}, D.~A. 2008, \pasp, 120, 531

\bibitem[{{Cushing} {et~al.}(2005){Cushing}, {Rayner}, \&
  {Vacca}}]{Cushing:2005lr}
{Cushing}, M.~C., {Rayner}, J.~T., \& {Vacca}, W.~D. 2005, \apj, 623, 1115

\bibitem[{{Cushing} {et~al.}(2004){Cushing}, {Vacca}, \&
  {Rayner}}]{Cushing:2004fk}
{Cushing}, M.~C., {Vacca}, W.~D., \& {Rayner}, J.~T. 2004, \pasp, 116, 362

\bibitem[{{de Bruijne}(2012)}]{deBruijne2012}
{de Bruijne}, J.~H.~J. 2012, \apss, 341, 31

\bibitem[{{Delfosse} {et~al.}(2000){Delfosse}, {Forveille}, {S{\'e}gransan},
  {Beuzit}, {Udry}, {Perrier}, \& {Mayor}}]{Delfosse2000}
{Delfosse}, X., {Forveille}, T., {S{\'e}gransan}, D., {Beuzit}, J.-L., {Udry},
  S., {Perrier}, C., \& {Mayor}, M. 2000, \aap, 364, 217

\bibitem[{{Dotter} {et~al.}(2008){Dotter}, {Chaboyer}, {Jevremovi{\'c}},
  {Kostov}, {Baron}, \& {Ferguson}}]{Dotter2008}
{Dotter}, A., {Chaboyer}, B., {Jevremovi{\'c}}, D., {Kostov}, V., {Baron}, E.,
  \& {Ferguson}, J.~W. 2008, \apjs, 178, 89

\bibitem[{{Fischer} \& {Valenti}(2005)}]{Fischer2005}
{Fischer}, D.~A., \& {Valenti}, J. 2005, \apj, 622, 1102

\bibitem[{{Gaidos}(2013)}]{Gaidos2013a}
{Gaidos}, E. 2013, \apj, 770, 90

\bibitem[{{Gaidos} {et~al.}(2013){Gaidos}, {Fischer}, {Mann}, \&
  {Howard}}]{Gaidos2013b}
{Gaidos}, E., {Fischer}, D.~A., {Mann}, A.~W., \& {Howard}, A.~W. 2013, \apj,
  771, 18

\bibitem[{{Gaidos} {et~al.}(2012){Gaidos}, {Fischer}, {Mann}, \&
  {L{\'e}pine}}]{Gaidos2012}
{Gaidos}, E., {Fischer}, D.~A., {Mann}, A.~W., \& {L{\'e}pine}, S. 2012, \apj,
  746, 36

\bibitem[{{Gonzalez}(1998)}]{Gonzalez1998}
{Gonzalez}, G. 1998, \aap, 334, 221

\bibitem[{{Hamuy} {et~al.}(1994){Hamuy}, {Suntzeff}, {Heathcote}, {Walker},
  {Gigoux}, \& {Phillips}}]{Hamuy1994}
{Hamuy}, M., {Suntzeff}, N.~B., {Heathcote}, S.~R., {Walker}, A.~R., {Gigoux},
  P., \& {Phillips}, M.~M. 1994, \pasp, 106, 566

\bibitem[{{Hasegawa} \& {Ida}(2013)}]{Hasegawa2013}
{Hasegawa}, Y., \& {Ida}, S. 2013, \apj, 774, 146

\bibitem[{{Helled} {et~al.}(2013){Helled}, {Bodenheimer}, {Podolak}, {Boley},
  {Meru}, {Nayakshin}, {Fortney}, {Mayer}, {Alibert}, \& {Boss}}]{Helled2014}
{Helled}, R., {et~al.} 2013, ArXiv e-prints

\bibitem[{{Henry} \& {McCarthy}(1993)}]{Henry1993}
{Henry}, T.~J., \& {McCarthy}, Jr., D.~W. 1993, \aj, 106, 773

\bibitem[{{Howell} {et~al.}(2014){Howell}, {Sobeck}, {Haas}, {Still},
  {Barclay}, {Mullally}, {Troeltzsch}, {Aigrain}, {Bryson}, {Caldwell},
  {Chaplin}, {Cochran}, {Huber}, {Marcy}, {Miglio}, {Najita}, {Smith},
  {Twicken}, \& {Fortney}}]{Howell2014}
{Howell}, S.~B., {et~al.} 2014, \pasp, 126, 398

\bibitem[{{Johnson} {et~al.}(2010){Johnson}, {Aller}, {Howard}, \&
  {Crepp}}]{Johnson2010}
{Johnson}, J.~A., {Aller}, K.~M., {Howard}, A.~W., \& {Crepp}, J.~R. 2010,
  \pasp, 122, 905

\bibitem[{{Johnson} \& {Apps}(2009)}]{Johnson2009}
{Johnson}, J.~A., \& {Apps}, K. 2009, \apj, 699, 933

\bibitem[{{Johnson} {et~al.}(2013){Johnson}, {Morton}, \&
  {Wright}}]{Johnson2013}
{Johnson}, J.~A., {Morton}, T.~D., \& {Wright}, J.~T. 2013, \apj, 763, 53

\bibitem[{{Kang} {et~al.}(2011){Kang}, {Lee}, \& {Kim}}]{Kang2011}
{Kang}, W., {Lee}, S.-G., \& {Kim}, K.-M. 2011, \apj, 736, 87

\bibitem[{{Lantz} {et~al.}(2004){Lantz}, {Aldering}, {Antilogus}, {Bonnaud},
  {Capoani}, {Castera}, {Copin}, {Dubet}, {Gangler}, {Henault}, {Lemonnier},
  {Pain}, {Pecontal}, {Pecontal}, \& {Smadja}}]{Lantz:2004}
{Lantz}, B., {et~al.} 2004, in Society of Photo-Optical Instrumentation
  Engineers (SPIE) Conference Series, Vol. 5249, Society of Photo-Optical
  Instrumentation Engineers (SPIE) Conference Series, ed. {L.~Mazuray,
  P.~J.~Rogers, \& R.~Wartmann}, 146--155

\bibitem[{{L{\'e}pine} \& {Gaidos}(2011)}]{Lepine:2011vn}
{L{\'e}pine}, S., \& {Gaidos}, E. 2011, \aj, 142, 138

\bibitem[{{L{\'e}pine} {et~al.}(2013){L{\'e}pine}, {Hilton}, {Mann}, {Wilde},
  {Rojas-Ayala}, {Cruz}, \& {Gaidos}}]{Lepine:2013}
{L{\'e}pine}, S., {Hilton}, E.~J., {Mann}, A.~W., {Wilde}, M., {Rojas-Ayala},
  B., {Cruz}, K.~L., \& {Gaidos}, E. 2013, \aj, 145, 102

\bibitem[{{Lloyd}(2011)}]{Lloyd2011}
{Lloyd}, J.~P. 2011, \apjl, 739, L49

\bibitem[{{Lloyd}(2013)}]{Lloyd2013}
---. 2013, \apjl, 774, L2

\bibitem[{{Mahadevan} {et~al.}(2012){Mahadevan}, {Ramsey}, {Bender}, {Terrien},
  {Wright}, {Halverson}, {Hearty}, {Nelson}, {Burton}, {Redman}, {Osterman},
  {Diddams}, {Kasting}, {Endl}, \& {Deshpande}}]{Mahadevan2012}
{Mahadevan}, S., {et~al.} 2012, in Society of Photo-Optical Instrumentation
  Engineers (SPIE) Conference Series, Vol. 8446, Society of Photo-Optical
  Instrumentation Engineers (SPIE) Conference Series

\bibitem[{{Mann} {et~al.}(2013{\natexlab{a}}){Mann}, {Brewer}, {Gaidos},
  {L{\'e}pine}, \& {Hilton}}]{Mann2013}
{Mann}, A.~W., {Brewer}, J.~M., {Gaidos}, E., {L{\'e}pine}, S., \& {Hilton},
  E.~J. 2013{\natexlab{a}}, \aj, 145, 52

\bibitem[{{Mann} {et~al.}(2013{\natexlab{b}}){Mann}, {Gaidos}, \&
  {Ansdell}}]{Mann2013c}
{Mann}, A.~W., {Gaidos}, E., \& {Ansdell}, M. 2013{\natexlab{b}}, \apj, 779,
  188

\bibitem[{{Mann} {et~al.}(2013{\natexlab{c}}){Mann}, {Gaidos}, {Kraus}, \&
  {Hilton}}]{Mann2013b}
{Mann}, A.~W., {Gaidos}, E., {Kraus}, A., \& {Hilton}, E.~J.
  2013{\natexlab{c}}, \apj, 770, 43

\bibitem[{{Markwardt}(2009)}]{Markwardt2009}
{Markwardt}, C.~B. 2009, in ASP Conference Series, Vol. 411, Astronomical Data
  Analysis Software and Systems XVIII, ed. D.~A. {Bohlender}, D.~{Durand}, \&
  P.~{Dowler}, 251

\bibitem[{{Montet} {et~al.}(2014){Montet}, {Crepp}, {Johnson}, {Howard}, \&
  {Marcy}}]{Montet2014}
{Montet}, B.~T., {Crepp}, J.~R., {Johnson}, J.~A., {Howard}, A.~W., \& {Marcy},
  G.~W. 2014, \apj, 781, 28

\bibitem[{{Moorhead} {et~al.}(2011){Moorhead}, {Ford}, {Morehead}, {Rowe},
  {Borucki}, {Batalha}, {Bryson}, {Caldwell}, {Fabrycky}, {Gautier}, {Koch},
  {Holman}, {Jenkins}, {Li}, {Lissauer}, {Lucas}, {Marcy}, {Quinn}, {Quintana},
  {Ragozzine}, {Shporer}, {Still}, \& {Torres}}]{Moorhead2011a}
{Moorhead}, A.~V., {et~al.} 2011, \apjs, 197, 1

\bibitem[{{Neves} {et~al.}(2013){Neves}, {Bonfils}, {Santos}, {Delfosse},
  {Forveille}, {Allard}, \& {Udry}}]{Neves2013}
{Neves}, V., {Bonfils}, X., {Santos}, N.~C., {Delfosse}, X., {Forveille}, T.,
  {Allard}, F., \& {Udry}, S. 2013, \aap, 551, A36

\bibitem[{{Neves} {et~al.}(2012){Neves}, {Bonfils}, {Santos}, {Delfosse},
  {Forveille}, {Allard}, {Nat{\'a}rio}, {Fernandes}, \& {Udry}}]{Neves2012}
{Neves}, V., {et~al.} 2012, \aap, 538, A25

\bibitem[{{Oke}(1990)}]{Oke1990}
{Oke}, J.~B. 1990, \aj, 99, 1621

\bibitem[{{{\"O}nehag} {et~al.}(2012){{\"O}nehag}, {Heiter}, {Gustafsson},
  {Piskunov}, {Plez}, \& {Reiners}}]{Onehag2012}
{{\"O}nehag}, A., {Heiter}, U., {Gustafsson}, B., {Piskunov}, N., {Plez}, B.,
  \& {Reiners}, A. 2012, \aap, 542, A33

\bibitem[{{Owen} {et~al.}(2012){Owen}, {Clarke}, \& {Ercolano}}]{Owen2012}
{Owen}, J.~E., {Clarke}, C.~J., \& {Ercolano}, B. 2012, \mnras, 422, 1880

\bibitem[{{Pineda} {et~al.}(2013){Pineda}, {Bottom}, \& {Johnson}}]{Pineda2013}
{Pineda}, J.~S., {Bottom}, M., \& {Johnson}, J.~A. 2013, \apj, 767, 28

\bibitem[{{Quirrenbach} {et~al.}(2012){Quirrenbach}, {Amado}, {Seifert},
  {S{\'a}nchez Carrasco}, {Mandel}, {Caballero}, {Mundt}, {Ribas}, {Reiners},
  {Abril}, {Aceituno}, {Alonso-Floriano}, {Ammler-von Eiff}, {Anglada-Escude},
  {Antona Jim{\'e}nez}, {Anwand-Heerwart}, {Barrado y Navascu{\'e}s},
  {Becerril}, {Bejar}, {Benitez}, {Cardenas}, {Claret}, {Colome},
  {Cort{\'e}s-Contreras}, {Czesla}, {del Burgo}, {Doellinger}, {Dorda},
  {Dreizler}, {Feiz}, {Fernandez}, {Galadi}, {Garrido}, {Gonz{\'a}lez
  Hern{\'a}ndez}, {Guardia}, {Guenther}, {de Guindos}, {Guti{\'e}rrez-Soto},
  {Hagen}, {Hatzes}, {Hauschildt}, {Helmling}, {Henning}, {Herrero}, {Huber},
  {Huber}, {Jeffers}, {Joergens}, {de Juan}, {Kehr}, {Klutsch}, {K{\"u}rster},
  {Lalitha}, {Laun}, {Lemke}, {Lenzen}, {Lizon}, {L{\'o}pez del Fresno},
  {L{\'o}pez-Morales}, {L{\'o}pez-Santiago}, {Mall}, {Martin},
  {Mart{\'{\i}}n-Ruiz}, {Mirabet}, {Montes}, {Morales}, {Morales Mu{\~n}oz},
  {Moya}, {Naranjo}, {Oreiro}, {P{\'e}rez Medialdea}, {Pluto}, {Rabaza},
  {Ramon}, {Rebolo}, {Reffert}, {Rhode}, {Rix}, {Rodler}, {Rodr{\'{\i}}guez},
  {Rodr{\'{\i}}guez L{\'o}pez}, {Rodr{\'{\i}}guez P{\'e}rez}, {Rodriguez
  Trinidad}, {Rohloff}, {S{\'a}nchez-Blanco}, {Sanz-Forcada}, {Sch{\"a}fer},
  {Schiller}, {Schmidt}, {Schmitt}, {Solano}, {Stahl}, {Storz}, {St{\"u}rmer},
  {Suarez}, {Thiele}, {Ulbrich}, {Vidal-Dasilva}, {Wagner}, {Winkler}, {Xu},
  {Zapatero Osorio}, \& {Zechmeister}}]{Quirrenbach2012}
{Quirrenbach}, A., {et~al.} 2012, in Society of Photo-Optical Instrumentation
  Engineers (SPIE) Conference Series, Vol. 8446, Society of Photo-Optical
  Instrumentation Engineers (SPIE) Conference Series

\bibitem[{{Rajpurohit} {et~al.}(2014){Rajpurohit}, {Reyle}, {Allard}, {Scholz},
  {Homeier}, {Schultheis}, \& {Bayo}}]{Rajpurohit2014}
{Rajpurohit}, A.~S., {Reyle}, C., {Allard}, F., {Scholz}, R.-D., {Homeier}, D.,
  {Schultheis}, M., \& {Bayo}, A. 2014, ArXiv e-prints

\bibitem[{{Rauscher} \& {Marcy}(2006)}]{Rauscher2006}
{Rauscher}, E., \& {Marcy}, G.~W. 2006, \pasp, 118, 617

\bibitem[{{Rayner} {et~al.}(2009){Rayner}, {Cushing}, \&
  {Vacca}}]{Rayner:2009kx}
{Rayner}, J.~T., {Cushing}, M.~C., \& {Vacca}, W.~D. 2009, \apjs, 185, 289

\bibitem[{{Rayner} {et~al.}(2003){Rayner}, {Toomey}, {Onaka}, {Denault},
  {Stahlberger}, {Vacca}, {Cushing}, \& {Wang}}]{Rayner:2003lr}
{Rayner}, J.~T., {Toomey}, D.~W., {Onaka}, P.~M., {Denault}, A.~J.,
  {Stahlberger}, W.~E., {Vacca}, W.~D., {Cushing}, M.~C., \& {Wang}, S. 2003,
  \pasp, 115, 362

\bibitem[{{Ricker} {et~al.}(2014){Ricker}, {Winn}, {Vanderspek}, {Latham},
  {Bakos}, {Bean}, {Berta-Thompson}, {Brown}, {Buchhave}, {Butler}, {Butler},
  {Chaplin}, {Charbonneau}, {Christensen-Dalsgaard}, {Clampin}, {Deming},
  {Doty}, {De Lee}, {Dressing}, {Dunham}, {Endl}, {Fressin}, {Ge}, {Henning},
  {Holman}, {Howard}, {Ida}, {Jenkins}, {Jernigan}, {Johnson}, {Kaltenegger},
  {Kawai}, {Kjeldsen}, {Laughlin}, {Levine}, {Lin}, {Lissauer}, {MacQueen},
  {Marcy}, {McCullough}, {Morton}, {Narita}, {Paegert}, {Palle}, {Pepe},
  {Pepper}, {Quirrenbach}, {Rinehart}, {Sasselov}, {Sato}, {Seager},
  {Sozzetti}, {Stassun}, {Sullivan}, {Szentgyorgyi}, {Torres}, {Udry}, \&
  {Villasenor}}]{Ricker2014}
{Ricker}, G.~R., {et~al.} 2014, ArXiv e-prints

\bibitem[{Rojas-Ayala {et~al.}(2012)Rojas-Ayala, Covey, Muirhead, \&
  Lloyd}]{RojasAyala2012}
Rojas-Ayala, B., Covey, K., Muirhead, P., \& Lloyd, J.~P. 2012, Astrophys. J.,
  748, 93

\bibitem[{{Rojas-Ayala} {et~al.}(2010){Rojas-Ayala}, {Covey}, {Muirhead}, \&
  {Lloyd}}]{Rojas-Ayala2010}
{Rojas-Ayala}, B., {Covey}, K.~R., {Muirhead}, P.~S., \& {Lloyd}, J.~P. 2010,
  \apjl, 720, L113

\bibitem[{{Rojas-Ayala} {et~al.}(2012){Rojas-Ayala}, {Covey}, {Muirhead}, \&
  {Lloyd}}]{Rojas-Ayala2012}
---. 2012, \apj, 748, 93

\bibitem[{{Santos} {et~al.}(2004){Santos}, {Israelian}, \&
  {Mayor}}]{Santos2004}
{Santos}, N.~C., {Israelian}, G., \& {Mayor}, M. 2004, \aap, 415, 1153

\bibitem[{{Santos} {et~al.}(2003){Santos}, {Israelian}, {Mayor}, {Rebolo}, \&
  {Udry}}]{Santos2003}
{Santos}, N.~C., {Israelian}, G., {Mayor}, M., {Rebolo}, R., \& {Udry}, S.
  2003, \aap, 398, 363

\bibitem[{{Schlaufman} \& {Laughlin}(2010)}]{Schlaufman2010}
{Schlaufman}, K.~C., \& {Laughlin}, G. 2010, \aap, 519, A105

\bibitem[{{Schlaufman} \& {Winn}(2013)}]{Schlaufman2013}
{Schlaufman}, K.~C., \& {Winn}, J.~N. 2013, \apj, 772, 143

\bibitem[{{Sousa} {et~al.}(2008){Sousa}, {Santos}, {Mayor}, {Udry},
  {Casagrande}, {Israelian}, {Pepe}, {Queloz}, \& {Monteiro}}]{Sousa2008}
{Sousa}, S.~G., {et~al.} 2008, \aap, 487, 373

\bibitem[{{Sozzetti} {et~al.}(2014){Sozzetti}, {Giacobbe}, {Lattanzi},
  {Micela}, {Morbidelli}, \& {Tinetti}}]{Sozzetti2014}
{Sozzetti}, A., {Giacobbe}, P., {Lattanzi}, M.~G., {Micela}, G., {Morbidelli},
  R., \& {Tinetti}, G. 2014, \mnras, 437, 497

\bibitem[{{Tamura} {et~al.}(2012){Tamura}, {Suto}, {Nishikawa}, {Kotani},
  {Sato}, {Aoki}, {Usuda}, {Kurokawa}, {Kashiwagi}, {Nishiyama}, {Ikeda},
  {Hall}, {Hodapp}, {Hashimoto}, {Morino}, {Inoue}, {Mizuno}, {Washizaki},
  {Tanaka}, {Suzuki}, {Kwon}, {Suenaga}, {Oh}, {Narita}, {Kokubo}, {Hayano},
  {Izumiura}, {Kambe}, {Kudo}, {Kusakabe}, {Ikoma}, {Hori}, {Omiya}, {Genda},
  {Fukui}, {Fujii}, {Guyon}, {Harakawa}, {Hayashi}, {Hidai}, {Hirano},
  {Kuzuhara}, {Machida}, {Matsuo}, {Nagata}, {Ohnuki}, {Ogihara}, {Oshino},
  {Suzuki}, {Takami}, {Takato}, {Takahashi}, {Tachinami}, \&
  {Terada}}]{Tamura2012}
{Tamura}, M., {et~al.} 2012, in Society of Photo-Optical Instrumentation
  Engineers (SPIE) Conference Series, Vol. 8446, Society of Photo-Optical
  Instrumentation Engineers (SPIE) Conference Series

\bibitem[{{Terrien} {et~al.}(2012){Terrien}, {Mahadevan}, {Bender},
  {Deshpande}, {Ramsey}, \& {Bochanski}}]{Terrien2012}
{Terrien}, R.~C., {Mahadevan}, S., {Bender}, C.~F., {Deshpande}, R., {Ramsey},
  L.~W., \& {Bochanski}, J.~J. 2012, \apjl, 747, L38

\bibitem[{{Thibault} {et~al.}(2012){Thibault}, {Rabou}, {Donati},
  {Desaulniers}, {Dallaire}, {Artigau}, {Pepe}, {Micheau}, {Vall{\'e}e},
  {Pepe}, {Barrick}, {Reshetov}, {Hernandez}, {Saddlemyer}, {Pazder},
  {Par{\`e}s}, {Doyon}, {Delfosse}, {Kouach}, \& {Loop}}]{Thibault2012}
{Thibault}, S., {et~al.} 2012, in Society of Photo-Optical Instrumentation
  Engineers (SPIE) Conference Series, Vol. 8446, Society of Photo-Optical
  Instrumentation Engineers (SPIE) Conference Series

\bibitem[{{Torres} {et~al.}(2012){Torres}, {Fischer}, {Sozzetti}, {Buchhave},
  {Winn}, {Holman}, \& {Carter}}]{Torres2012}
{Torres}, G., {Fischer}, D.~A., {Sozzetti}, A., {Buchhave}, L.~A., {Winn},
  J.~N., {Holman}, M.~J., \& {Carter}, J.~A. 2012, \apj, 757, 161

\bibitem[{{Vacca} {et~al.}(2003){Vacca}, {Cushing}, \& {Rayner}}]{Vacca:2003qy}
{Vacca}, W.~D., {Cushing}, M.~C., \& {Rayner}, J.~T. 2003, \pasp, 115, 389

\bibitem[{{Valenti} \& {Fischer}(2005)}]{Valenti2005}
{Valenti}, J.~A., \& {Fischer}, D.~A. 2005, \apjs, 159, 141

\bibitem[{{Wright} {et~al.}(2011{\natexlab{a}}){Wright}, {Veras}, {Ford},
  {Johnson}, {Marcy}, {Howard}, {Isaacson}, {Fischer}, {Spronck}, {Anderson},
  \& {Valenti}}]{Wright2011b}
{Wright}, J.~T., {et~al.} 2011{\natexlab{a}}, \apj, 730, 93

\bibitem[{{Wright} {et~al.}(2011{\natexlab{b}}){Wright}, {Fakhouri}, {Marcy},
  {Han}, {Feng}, {Johnson}, {Howard}, {Fischer}, {Valenti}, {Anderson}, \&
  {Piskunov}}]{Wright2011}
---. 2011{\natexlab{b}}, \pasp, 123, 412

\bibitem[{{Yasui} {et~al.}(2009){Yasui}, {Kobayashi}, {Tokunaga}, {Saito}, \&
  {Tokoku}}]{Yasui2009}
{Yasui}, C., {Kobayashi}, N., {Tokunaga}, A.~T., {Saito}, M., \& {Tokoku}, C.
  2009, \apj, 705, 54

\end{thebibliography}
\end{document}